# Length-dependent thermal conductivity in suspended single layer graphene

Xiangfan Xu[1,2,3,†], Luiz F. C. Pereira[4,†], Yu Wang[5], Jing Wu[1,2], Kaiwen Zhang[1,2,6], Xiangming Zhao[1,2,6], Sukang Bae[7], Cong Tinh Bui[8], Rongguo Xie[1,6,9], John T. L. Thong[9,8], Byung Hee Hong[10], Kian Ping Loh[2,8,11], Davide Donadio[4], Baowen Li[1,2,6,8], Barbaros Özyilmaz[1,2,3,8]

[1] Department of Physics, National University of Singapore, Singapore 117542

[2] Graphene Research Center, National University of Singapore, Singapore 117542

[3] NanoCore, 4 Engineering Drive 3, National University of Singapore, Singapore 117576

[4] Max Planck Institute for Polymer Research, Ackermannweg 10, 55128 Mainz, Germany

[5] State Key Laboratory of Multiphase Complex Systems, Institute of Process Engineering, Chinese Academy of Science, Beijing 100190, P. R. China

[6] Centre for Computational Science and Engineering, National University of Singapore, Singapore 117542

[7] SKKU Advanced Institute of Nanotechnology (SAINT) and Center for Human Interface Nano Technology (HINT), Sungkyunkwan University, Suwon 440-746, Republic of Korea

[8] NUS Graduate School for Integrative Science and Engineering, Singapore 117456

[9] Department of Electrical and Computer Engineering, National University of Singapore, Singapore 117576

[10] Department of Chemistry, Seoul National University, Seoul 152-742, Republic of Korea

[11] Department of Chemistry, National University of Singapore, Singapore 117543

† Present address: Center for Phononics and Thermal Energy Science, School of Physics Science and Engineering, Tongji University, Shanghai 200092, P. R. China (X.X.); Departamento de Física Teórica e Experimental, Universidade Federal do Rio Grande do Norte, 59078-900, Natal-RN, Brazil (L.F.C.P.).

Correspondence and requests for materials should be addressed to B. Ö. (email: barbaros@nus.edu.sg) or to B. L. (phylibw@nus.edu.sg) or to D. D. (email: donadio@mpip-mainz.mpg.de).

**Graphene exhibits extraordinary electronic and mechanical properties, and extremely high thermal conductivity. Being a very stable atomically thick membrane that can be suspended between two leads, graphene provides a perfect test platform for studying thermal conductivity in two-dimensional systems, which is of primary importance for phonon transport in low-dimensional materials. Here we report experimental measurements and non-equilibrium molecular dynamics simulations of thermal conduction in suspended single layer graphene as a function of both temperature and sample length. Interestingly and in contrast to bulk materials, when temperature at 300K, thermal conductivity keeps increasing and remains logarithmic divergence with sample length even for sample lengths much larger than the average phonon mean free path. This result is a consequence of the two-dimensional nature of phonons in graphene and provides fundamental understanding into thermal transport in two-dimensional materials.**

In bulk materials, heat conduction is governed by Fourier's law: $J = -\kappa \nabla T$, where $J$ is the local heat flux, $\nabla T$ is the temperature gradient, $\kappa$ is thermal conductivity which is an intrinsic parameter of the materials and is usually size and geometry independent. Over the past 200 years, Fourier's law has been successfully employed to describe thermal conduction in three-dimensional (3D) systems. However, with the recent availability of low-dimensional materials it remains to verify whether thermal conductivity is still size/geometry independent. For one-dimensional (1D) lattice models [1-5] and quasi-1D nanostructures and polymers [6-8] a length dependence of $\kappa$ as ~ $L^\beta$ ($L$ is sample length) has been predicted theoretically and then experimentally verified for nanotubes[9]. The same question remains open for two-dimensional (2D) systems. While logarithmic divergence of the thermal conductivity has been demonstrated numerically for 2D lattice models [2,3,10-13], simulations of bi-dimensional materials such as graphene, give contradictory results, which depend on the employed simulation techniques and approximations [14-17]. Graphene is believed to be a perfect test bed to investigate this size effect of 2D systems due to its novel thermal properties [18-33]. However, besides a recent experiment carried out on supported graphene nanoribbons[34], in which size effect, especially the length effect, is affected by graphene-substrate interaction[35], a compelling experimental investigation of length-dependent thermal conductivity in suspended single layer graphene (SLG) is still missing, probably due to the challenges in suspending this atomically thick membrane suitable for size-dependent thermal measurements.

Here we report measurements of thermal conduction in suspended SLG grown by chemical vapor deposition on copper (Cu-CVD)[36-38]. Both the experimental and simulation results show that thermal conductivity scale with length as $\kappa \sim \log L$ at room temperature even when $L$ is one order of magnitude larger than the average phonon mean free path (MFP), which we think is related to the 2D nature of phonons in graphene. Possible mechanisms of this divergent behavior are discussed.

**Results**

**Temperature-dependent thermal conduction.** The experimentally measured thermal conductance $\sigma = 1/R_{total}$ and $\kappa$ for all junctions are shown in Figure 1c&d. Here $\kappa$ is obtained from $\kappa = \frac{L}{A(R_{total} - R_c)}$, in which $R_{total}$ is the total measured thermal resistance, $R_c$ is the thermal contact resistance, and $A$ is the unit cross section area (we define $A = w \times h$, where $w$ is the width of the samples and $h \approx 0.34$ nm is the nominal thickness of single layer graphene). Independently on the sample length, all samples exhibit qualitatively the same behavior as a function of the temperature: $\sigma(T)$ increases with temperature and reaches a broad plateau. The thermal conductivity at $T = 300$ K in our longest sample ($L = 9$ μm) shows a value of $(1689 \pm 100)$ Wm$^{-1}$K$^{-1}$ ~ $(1813 \pm 111)$ Wm$^{-1}$K$^{-1}$ (see Supplementary Note 1 and Supplementary Table 1&2 for the error bars), which is comparable to the values obtained from Raman based measurement techniques (from ~1800 Wm$^{-1}$K$^{-1}$ to ~5300 Wm$^{-1}$K$^{-1}$, see Supplementary Note 2 and Supplementary Table 3). Remarkably, our measurements show that $\kappa$ increases with sample length over the entire measured temperature range.

**Quasi-ballistic phonon transport at low temperature in submicron samples.** Here plotting $\sigma/A$ is more instructive than $\kappa$. In all submicron samples the thermal conductance exhibits the same temperature dependence (Figure 2a). This has important implications for the nature of 2D phonon transport. It has been reported that phonons in graphene can travel without scattering, i.e. ballistic transport [34,39]. In clean devices, $\sigma/A$ has been expected to follow $\sigma_{ballistic}/A \approx \left[\frac{1}{4.4\times10^5 T^{1.68}} + 1/(1.2\times10^{10})\right]^{-1}$ Wm$^{-2}$K$^{-1}$, as shown by the black dashed curve in the same figure [34]. Indeed, the experimentally measured values are not only length independent, but also within 39.8% of the predicted ballistic thermal conductance at $T$ = 30 K (phonons in quasi-ballistic regime), indicating high sample quality and clean surface in the measured graphene.

Atomistic simulations provide a powerful tool for validation and interpretation of experimental measurements, with the advantage that the limit of perfectly crystalline samples can be probed. We have performed non-equilibrium molecular dynamics (NEMD) [40] simulations of heat transport in graphene at similar conditions as the experiments. We considered periodic models of defect-free, isotopically pure and suspended graphene patches with size between 5.5 nm and 1.4 μm, at 300 K and at 1000 K. Our simulations show that at 300 K the thermal conductance remains constant for length (between the hot and the cold reservoir) up to ~ 80 nm ($\sigma/\sigma_0 = 0.84$ at ~ 80 nm, where $\sigma_0$ is the calculated thermal conductance in a 5.5 nm-long sample), indicating ballistic heat transport in this size range (Figure 2b).

**Length-dependent thermal conductivity.** Before we discuss the length dependence of thermal transport at room temperature, we need to comment on the issue of thermal contact resistance $R_c$, a critical parameter in thermal conductivity studies in general. Note that while achieving low values for $R_c$ with 1D nanowires and nanotubes remains difficult, the inherently larger contact area with 2D sheets makes it less challenging. Our calculations give a relatively small value of $R_c$ with values ranging from ~ $9.9 \times 10^4$ K/W to ~ $1.26 \times 10^6$ K/W, which contributes ~ 0.9% to ~ 11.5% of $R_{total}$ in the sample with 9 μm length (see Supplementary Note 3 and Supplementary Table 4).

Therefore, we plot $\kappa$ vs. $L$ in Figure 3a for different thermal contact resistance, i.e. negligible $R_c$, $R_c = 5.5 \times 10^5$ K/W (~ 5% of $R_{total}$ in 9 μm sample), and $R_c = 1.26 \times 10^6$ K/W (~ 11.5% of $R_{total}$ in 9 μm sample), respectively. $\kappa$ increases with sample length at $T = 300$K. Surprisingly and in contrast with experiments on supported graphene[34], $\kappa$ does not saturate even in our longest sample with $L = 9$ μm. More importantly, $\kappa$ scales with ~ log$L$ in samples with $L \geq 700$ nm (we do not claim the ~ log$L$ behavior in shorter samples due to the large effects from $R_c$). Such behavior has been predicted for 2D systems and is a direct consequence of the 2D nature of phonons[2,3]. This anomalous length-dependent thermal conductivity should not be confused with quasi-ballistic transport discussed above, which occurs when the length of the sample is comparable to the average phonon mean free path (MFP) of the main heat carriers, i.e. acoustic phonons in graphene. Following the parameters used by Ghosh *et al.*[28], we estimate the average MFP to be around 240 nm at $T = 300$ K. Please note that MFP is a spectral property that depends on frequency and polarization of each phonon, therefore MFP = 240nm is an average or effective MFP of all the phonons contributing to thermal conduction. Moreover, our simulation results show that the thermal conductance starts decreasing for length between 43 nm to 80 nm (triangles in Figure 2b) and is $\sigma/\sigma_0$ ~ 0.6 for $L$ ~ 171 nm. Therefore, $L >$ MFP holds in our shortest samples at $T = 300$ K and $\kappa$ does not saturate even when the length of the sample is more than thirty times larger than the average MFP.

It is worth pointing out that, in contrast with 2D lattice models[10,13], graphene atoms can move out of plane, giving rise to flexural acoustic (ZA) and flexural optical (ZO) modes. The role of the ZA modes of graphene in heat transport is still debated: according to lattice dynamics calculations and equilibrium MD simulations the ZA modes provide the fundamental scattering channels to limit $\kappa$ to a finite value, unless strain is applied [16,17]. On the other hand calculations that take into account the deviations of phonon populations from equilibrium, through a self-consistent solution of the Peierls-Boltzmann equation, suggest that ZA modes are the main heat carriers in graphene, and show a significant size dependence of $\kappa$ at the micrometer scale [41]. Here we have probed direct stationary non-equilibrium heat transport at conditions very similar to the experiments by NEMD simulations. In sharp contrast with previous equilibrium MD simulations, but in excellent agreement with experiments, we observe $\kappa(L) \sim \log L$ both at $T = 300$ K and at 1000 K, for systems with size well beyond the ballistic length (Figure 3b). Surprisingly, in spite of the unavoidable presence of boundary scattering at the interface with the thermal reservoirs, $\kappa$ from NEMD exceeds the converged value of $\kappa$ from equilibrium Green-Kubo calculations, computed with the same empirical potential, indicating that the mechanism of heat transport is substantially different at stationary non-equilibrium. We also note that different NEMD approaches, namely reverse NEMD [40] and direct NEMD [42], give the same result at any given size. In addition for each $L$, $\kappa(L)$ is not sensitive to the magnitude of the applied heat flux or temperature gradients. In our NEMD simulations we ensured that grad($T$) is proportional to $J$, meaning that we do not break the linear response regime (see Supplementary Note 4 and Supplementary Fig. 1). We have also carefully considered the effect of different aspect ratios of the simulation cell, probing very large systems of square aspect ratio, which give results in agreement with those obtained with elongated supercells (Figure 3b), and confirm that the log(*L*) divergence is not an artifact of the simulation setup.

**Discussion**

We now seek to understand the physics behind the observed length-dependent thermal conductivity. A number of theoretical and computational studies on ideal systems, like the Fermi-Pasta-Ulam model, have demonstrated that, $\kappa$ in low-dimensional systems depends on system size (insert of Figure 3a). In fact, rigorous mathematical proof tells us that for a momentum conserved 1D system $\kappa$ diverges with system size [1-4], which has been confirmed experimentally in nanotubes[9]. On the other hand, the length dependence of 2D systems remains unresolved and a ~ log*L* dependence has been proposed by various analytical theories [2,3,10,13]. Our experiment and simulation studies provide strong evidence for a ~ log*L* behavior in graphene, which is different from both of that in 1D and 3D systems.

The other possible mechanism behind the length-dependent thermal conductivity is related to the ballistic propagation of extremely long wavelength, low frequency acoustic phonons[29,43]. Nika *et al.*[29] emphasized the importance of low frequency acoustic phonons in graphene. As sample size increases, more low frequency acoustic phonons can be excited and contribute to thermal conduction, resulting in a length-dependent behavior. The third possibility is related to the selection rules for three-phonon scattering, the phase space of which is strongly restricted by the reduced dimensionality[41,44], however a non-logarithmic dependence of the thermal conductivity was reported when three-phonon process are considered to second order.

A comparative analysis of phonon populations in MD simulations at equilibrium and non-equilibrium conditions suggests that in the latter conditions the population of out-of-plane modes is augmented whereas in-plane modes with polarization in the direction of the heat flux propagation get slightly depopulated (see Supplementary Fig. 2). Such population imbalance produces the observed discrepancy between equilibrium and non-equilibrium simulations, promoting logarithmic divergence of $\kappa$ at stationary non-equilibrium (see Supplementary Note 4 and Methods for further details). Our simulations therefore suggest that logarithmic divergence stems from the combination of reduced dimensionality and displacements of phonon populations at stationary non-equilibrium conditions.

From experiments, it is still challenging to distinguish which transport mechanism dominates the observed length-dependent thermal conductivity. To this end, it would be constructive to show the relative contribution to thermal conductivity from phonons with different frequencies, or to study the thermal conductivity in samples with much larger size (e.g. 100 μm or even millimeter), which would however require the development of new measurement techniques.

In summary, we have studied thermal conductivity in suspended single layer graphene. Thermal conductivity has been observed to increase with the length of the samples and to scale as ~ log $L$, even when $L$ is one order of magnitude longer than the average phonon mean free path. Possible mechanisms have been discussed: MD simulations, in excellent agreement with experiments, confirm the divergence of $\kappa$ and suggest that it is related to the two-dimensional nature of phonons in graphene and to the change of the phonon population at stationary non-equilibrium conditions.

## Methods

**MD simulations.** Non-equilibrium molecular dynamics simulations were performed using LAMMPS (http://lammps.sandia.gov/). Interatomic forces were described by the Tersoff potential with a parameter set optimized for graphene[45]. Simulations employed periodic boundary conditions in the graphene plane, and each supercell was relaxed at the simulation temperature to achieve zero in-plane stress. Two different NEMD simulation methods were used, in which the thermal conductivity is calculated directly from the temperature gradient and the heat flux via Fourier's law. The simulation supercell is divided in contiguous slabs along the direction of heat propagation (see Supplementary Fig. 3). Each slab contains approximately 440 atoms and the temperature of each slab is calculated from its average kinetic energy ($K_E$) as $T = 2/3 \, K_E/k_B$. The two methods differ as for the mechanism employed to generate a stationary heat flux. In the direct NEMD method [42], independent Langevin thermostats were used to control the temperature of the cold and hot slabs. In the reverse non-equilibrium MD method [31] (RNEMD) the temperature gradient is imposed by swapping the velocity of the slowest molecule in the cold slab with the velocity of the fastest one in the hot slab. After a transient period, a stationary temperature gradient is established in the system due to the imposed heat flux. However, the velocity exchange disturbs the stability of the numerical integration of the equations of motion and introduces a drift in the total energy of the system. Velocities were exchanged every 200 steps (20 fs), and we used a timestep of 0.1 fs for a stable integration of the equations of motion. Each NEMD simulation ran for at least 40 million steps and the temperature gradient and heat flux were taken as the average over the final 10 million steps. Error bars are the standard deviation of the averaged data. Supplementary Fig. 4 shows the temperature profile for two supercell sizes, using both NEMD and RNEMD methods. Temperature gradients are taken as the average slope considering both sides of the simulation cell, only in the linear regions between cold and hot layers. Such temperature gradients, although larger in RNEMD simulations than in NEMD, are comparable to the ones in the experiment.

Initially, the simulation cells were increased in length while keeping a constant width of ~11 nm. This procedure implies an increase in the aspect ratio of the simulation cells from a ratio of 1 for the smallest cell (4,400 atoms), up to a ratio of ~ 250 for the longest cell (1.1 million atoms). In order to assure that the length dependence observed in our results was not related to the increase in aspect ratio, we also performed simulations with square cells up to ~ 560 nm (11.5 million atoms), which are in agreement with the observed log divergence as shown in Figure 3b.

Validation tests of the simulation methods are described in Supplementary Note 4 and Supplementary Fig. 1.

**Fabrication Details.** We employed typical pre-patterned heater wires [46,47] (see Figure 1a&b) to measure thermal conductance of suspended SLG grown by chemical vapor deposition on copper (Cu-CVD)[36,38]. Our device fabrication starts with $SiN_x$ membrane based heater structures (Supplementary Fig. 5a) optimized for length dependent studies. To obtain high quality samples, the CVD graphene was grown by re-using the copper catalyst [37], which in general results in better transport properties.

In order to decrease the thermal contact resistance, $SiN_x$ wafer and Pt pre-patterned electrodes were cleaned for an extended period of time (>10 minutes) in $O_2$ plasma, followed by immediate CVD graphene transfer. This helps to clean organic residue on the Pt electrodes and create suspended chemical bonds, resulting in a further improvement of the thermal contact to graphene[48]. The graphene sheets were then patterned by standard electron beam lithography (EBL), followed by an $O_2$ plasma step (see Supplementary Fig. 5c). In a second EBL step, 30 nm Cr/Au bars were deposited on both ends of the graphene stripes to ensure good thermal contact with the Pt electrodes underneath graphene (see Supplementary Fig. 5d). After suspending graphene by wet etching, the devices were dried using a critical point dryer to avoid damage due to surface tension. It is worth noting that we do not etch away all the silicon under $SiN_x$ pad (i.e. heater and sensor) but left few micrometers to few tens of micrometers thick silicon to act as a massive heat batch and make sure that heat spreads uniformly within the $SiN_x$ pad during thermal measurements. Graphene samples were annealed for 5 hours at 300 degree centigrade in flowing hydrogen ($H_2$: 150 ml/min, Ar: 150 ml/min, 1atm) to remove possible organic residues before measurements.

**Sample geometry.** The length dependence of thermal conductivity is studied by varying the channel length from 300 nm to 9 μm. Since the thermal conductivity is also weakly width (*w*) dependent when the width is larger than 1.5 μm (see Supplementary Note 5 and Supplementary Fig. 6), we have kept the width fixed at 1.5 μm. We define the direction of heat flow as length, and the perpendicular direction as width.

**Thermal conductivity measurement.** The suspended graphene sheet provides a thermal path between the two $SiN_x$ membranes that are otherwise thermally and electrically isolated from each other (Figure 1a). A µA-DC current combined with an AC current (100-200 nA) was applied to the heater resistor ($R_h$, red Pt coil in Figure 1a). The DC current was used to apply Joule heat in $R_h$ and to increase its temperature ($T_h$) from the environment temperature, $T_0$. The AC current was used to measure the resistance of $R_h$. The Joule heating in $R_h$ gradually dissipates through the six Pt/$SiN_x$ beams and the graphene sheet, which raises the temperature ($T_s$) in the sensor resistor ($R_s$, blue Pt coil in Figure 1a). In the steady state, the thermal conductance of the graphene sheet, $\sigma_G$, and each of the suspended Pt/$SiN_x$ beam, $\sigma_l$, can be obtained from

$$\sigma_l = \frac{Q_h + Q_l}{\Delta T_h + \Delta T_s} \quad .......(1)$$

$$\text{and} \quad \sigma_G = \frac{\sigma_l \Delta T_s}{\Delta T_h - \Delta T_s} \quad ......(2)$$

where $\Delta T_h$ and $\Delta T_s$ are the temperature rise in the $R_h$ and $R_s$, $Q_h$ and $Q_l$ are the Joule heat applied to the heater $R_h$ and the Pt leads. The temperature change in each membrane as function of applied Joule heat was kept in the linear range. The increase in temperature, $\Delta T_h$ and $\Delta T_s$, was controlled to below 5K to minimize thermal radiation between the two membranes. All measurements were performed under vacuum conditions better than $1 \times 10^{-6}$ mbar.

**Acknowledgements**

This work was supported in part by Singapore National Research Foundation Fellowship (NRF-RF2008-07), NRF-CRP grant (R-143-000-360-281), NRF-CRP grant (R-144-000-295-281), by MOE T2 Grant R-144-000-305-112, by A*STAR grant (R-143-000-360-311), by NUS NanoCore, and by the Singapore Millennium Foundation SMF/NUS Horizon grant (R-144-000-271-592/646). L.F.C.P. and D.D. acknowledge the provision of computational facilities and support by Rechenzentrum Garching of the Max Planck society (MPG), and access to the supercomputer JUGENE at the Jülich Supercomputing Centre under NIC project HMZ26. Financial support was provided by MPG under the MPRG program.

## Author Contributions

X.X. and B.Ö. designed the experiment, X.X. performed experiments, X.X., B.L. and B.Ö. analysed data and wrote the paper; L.F.C.P and D. D. designed and carried out the simulations and wrote the paper; Y.W., S.B., B.H. and K.L. grown CVD graphene; J.W., K.Z. and X.Z. helped prepare and measure samples; C.B., R.X. and J.T. prepared the prepatterned Pt/SiN$_x$ wafers; B.Ö. directed the project.

**Competing financial interests:** The authors declare no competing financial interests.

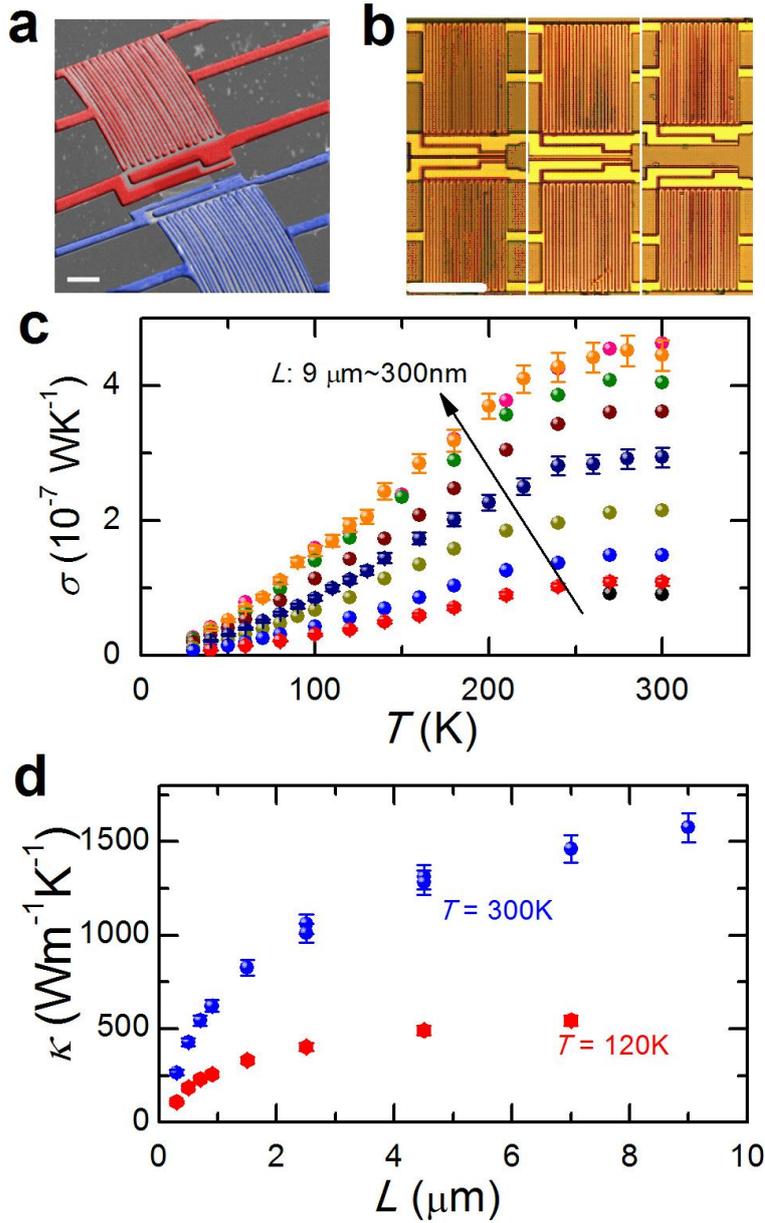

**Figure 1 | Thermal conduction vs. temperature.** (**a**) False-color scanning electron microscopy image of the suspended device, which consists of two 25 × 20 μm² Pt/SiN$_x$ membranes. The red and blue Pt coils are the heater ($R_h$) and sensor ($R_s$), which are thermally connected by suspended graphene (grey sheet in the middle). The scale bar is 5 μm. (**b**) SiN$_x$ membranes based heater structures optimized for length dependent studies. The scale bar is 20 μm. (**c**) Total measured thermal conductance ($\sigma$) vs. temperature. The sample with 9 μm broke when cooling below 270K during measurements. Also see Supplementary Fig. 7. (**d**) Thermal conductivity ($\kappa$) vs. sample length ($L$) with negligible thermal contact resistance $R_c$ at $T$ = 300K and 120K, respectively.

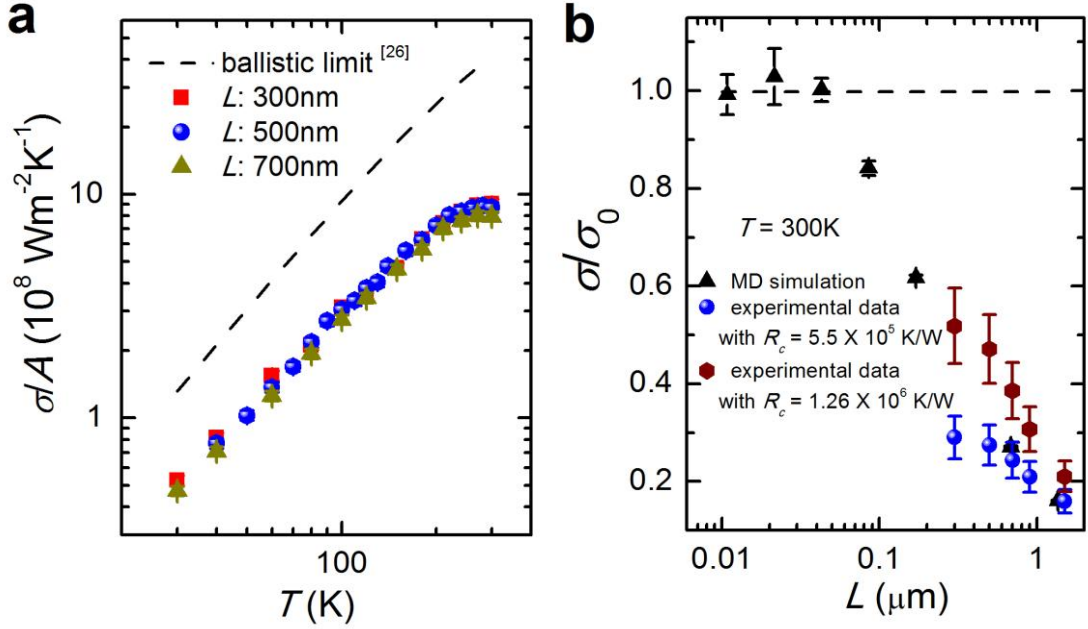

**Figure 2 | Thermal conductance in submicron devices**. (**a**) Experimental thermal conductance per unit cross section area $\sigma/A$ as a function of temperature in samples with $L$: 300 nm - 700 nm. The ballistic limit is indicated by the black dashed curve[34]. (**b**) Molecular dynamics (MD) simulation results (black triangles) of thermal conductance $\sigma/\sigma_0$ vs. sample length $L$ at $T = 300$ K. Simulation results are compared to experiments (blue circles and brown diamonds) assuming different values of thermal contact resistance $R_c$. The MD thermal conductance is normalized to $\sigma_0$, where $\sigma_0$ is the calculated thermal conductance in a sample with 5.5 nm between the hot and the cold reservoir; the experimental data are normalized the ballistic limit shown in Figure 2a at $T = 300$K, i.e. $\sigma_0/A = 4.17 \times 10^9$ Wm$^{-2}$K$^{-1}$.

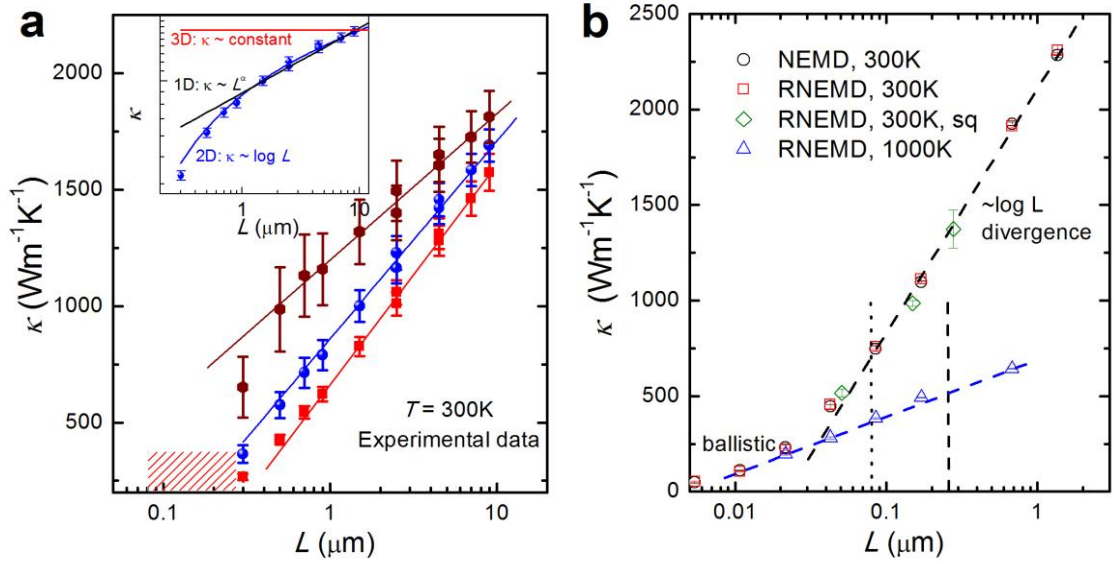

**Figure 3 | Experimental and simulation results on length-dependent thermal conductivity.** (**a**) Length dependence of the extracted intrinsic thermal conductivity when assuming $R_c$ contributes negligible (red squares), 5% (blue circles) and 11.5% (brown diamonds) to the total measured thermal resistance in 9μm-long sample. The linear solid lines are guides to the eyes. The red diagonal lines indicate the average/effective phonon mean free path MFP, which is 240 nm and 80nm obtained by experiment and simulation, respectively. Insert: illustration of log$\kappa$ ~ log$L$ scaling behavior for one-dimensional (1D), two-dimensional (2D) and three-dimensional (3D) systems, where thermal conductivity scales as ~ $L^{0.33}$, ~ log$L$ and constant, respectively. Also see Supplementary Fig. 8. (**b**)Thermal conductivity of graphene as a function of the distance between the hot and the cold reservoir, obtained with direct (black circles) and reverse non-equilibrium molecular dynamics NEMD (red squares and green diamonds) methods at $T$ = 300K and reverse NEMD at 1000 K (blue triangles). Results obtained with supercells with square aspect ratio (green diamonds) agree well with those obtained with elongated supercells (all others). Dashed lines are logarithmic fit. The dotted and dashed vertical lines indicate the limit of the ballistic transport regime obtained by experiment and simulation, respectively.

# Supplementary Information

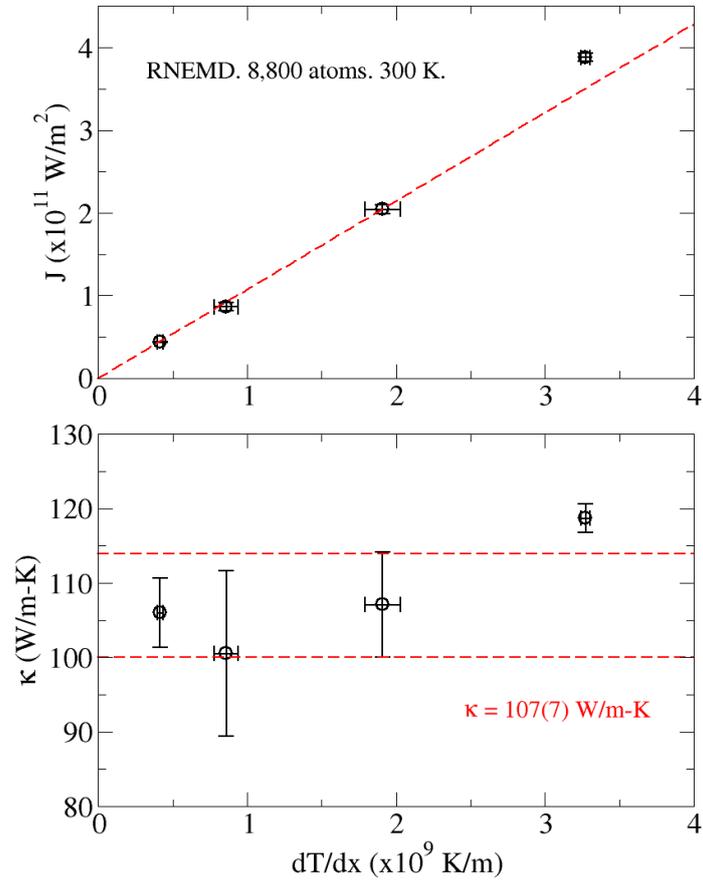

**Supplementary Figure 1 | Heat flux and thermal conductivity as a function of temperature gradient obtained from reverse NEMD simulations.**

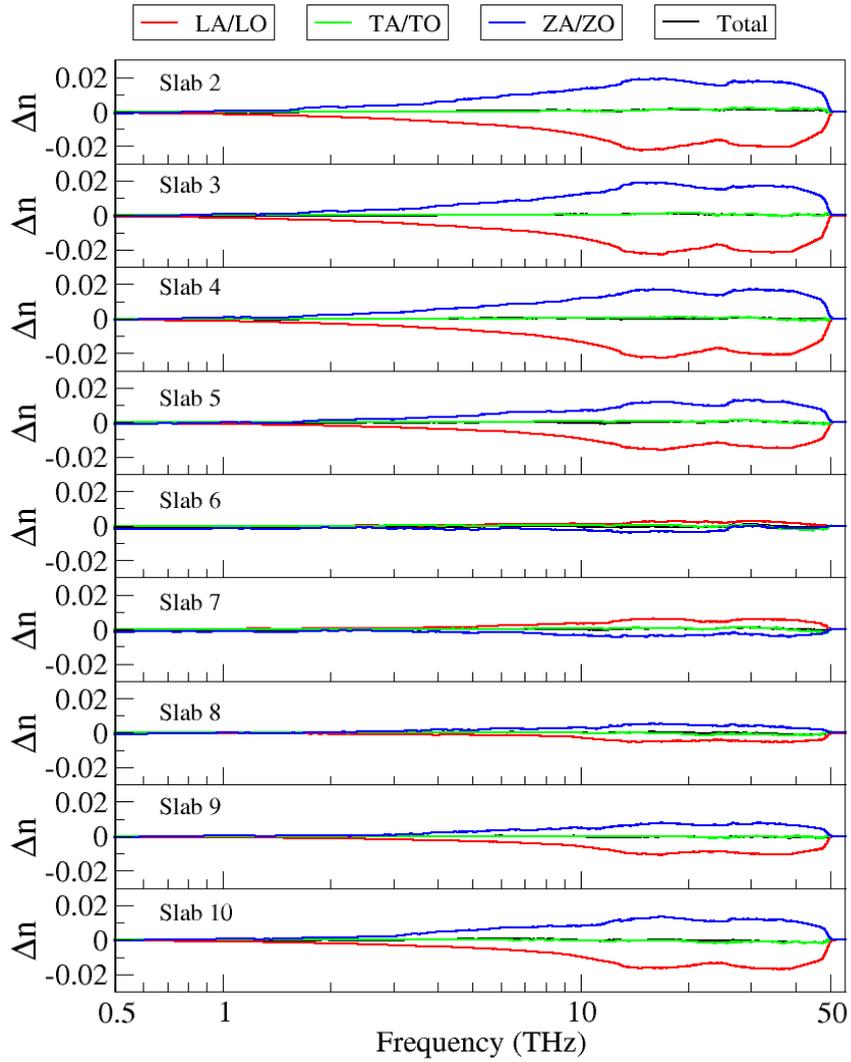

**Supplementary Figure 2 | Displacement of phonon populations as a function of frequency in non-equilibrium molecular dynamics simulations with respect to equilibrium populations for a system 170 nm long, i.e. in the ~ log(*L*) regime. Population displacements for phonons with different polarization are calculated for different slabs of the NEMD from the one next to the cold reservoir (slab 2) to the one near hot reservoir (slab 10).**

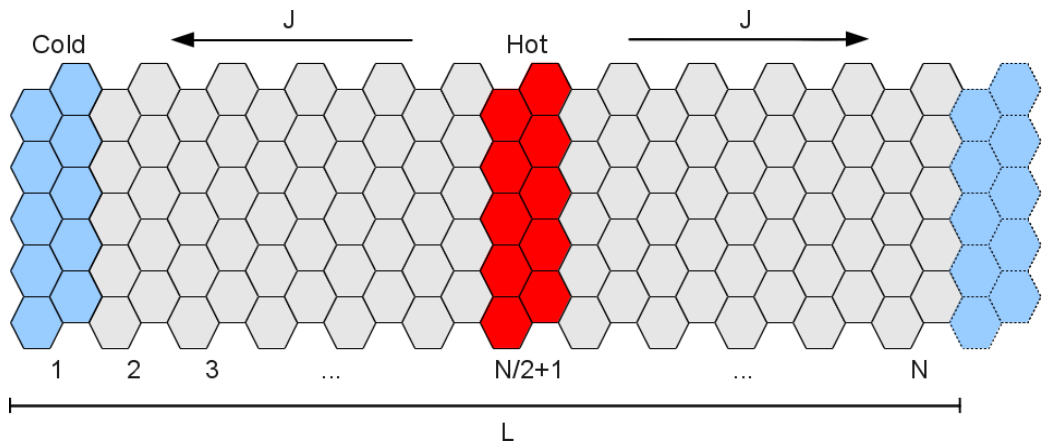

**Supplementary Figure 3 | Illustration of the simulation cell used in NEMD simulations**

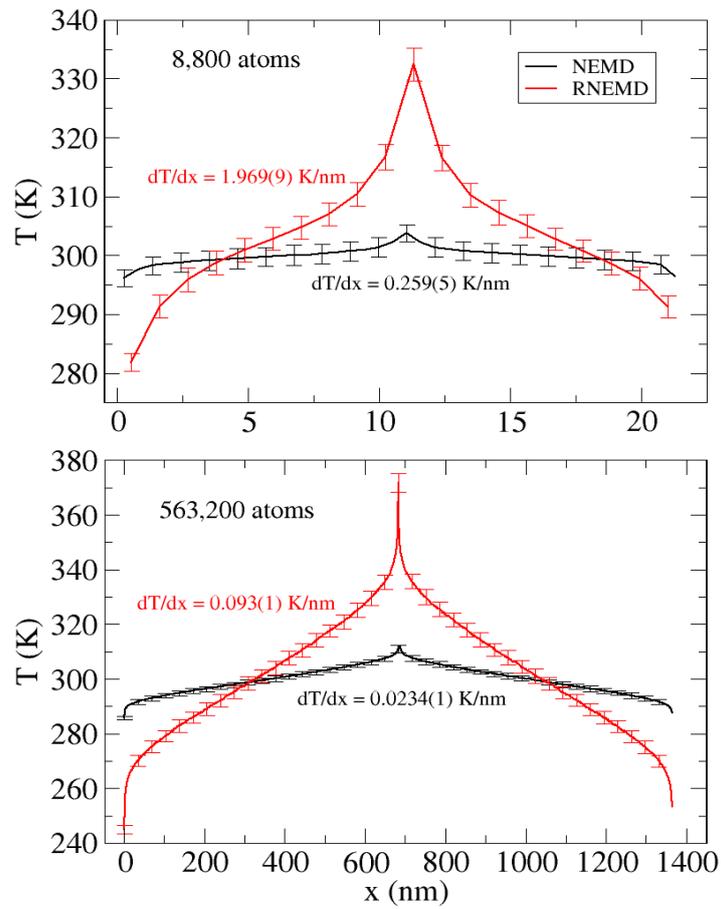

**Supplementary Figure 4 | Temperature profiles calculated from NEMD simulations for the second shortest and second longest simulation cells.**

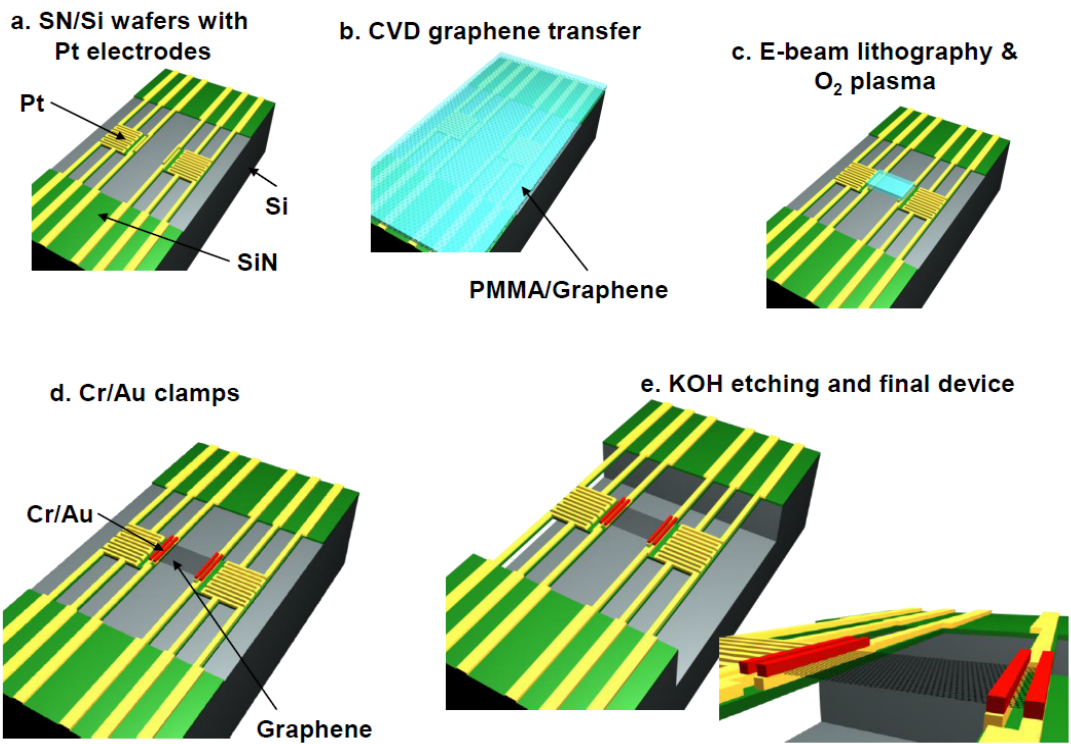

**Supplementary Figure 5 | Fabrication of single layer CVD graphene suitable for thermal conductivity measurements.**

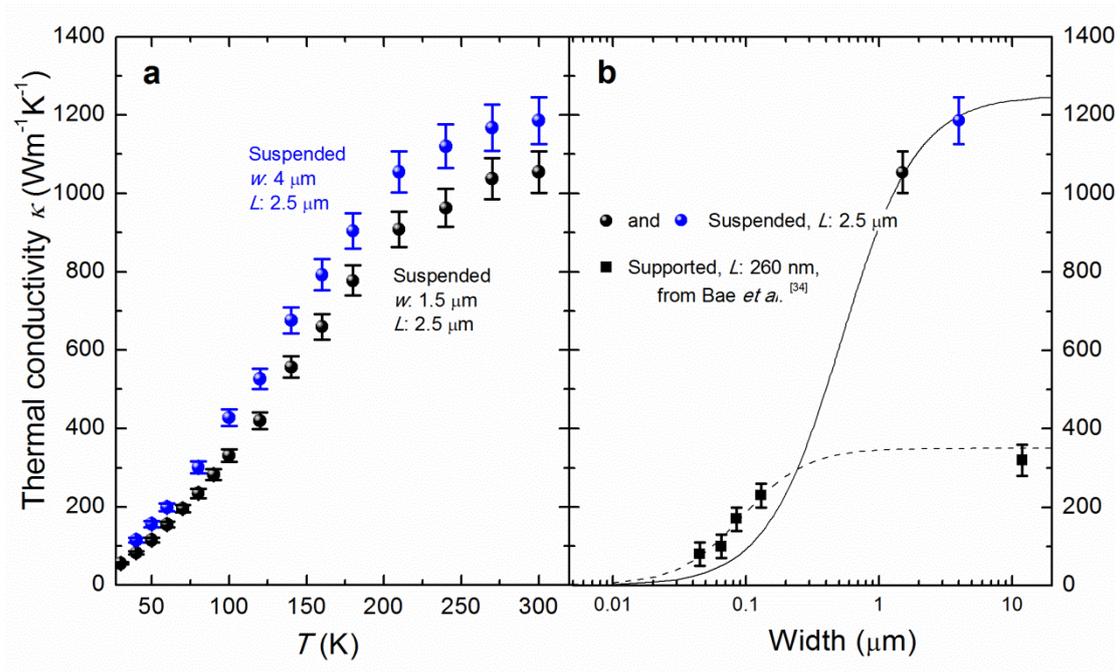

**Supplementary Figure 6 | Width dependent thermal conductivity.** (**a**) Temperature dependent thermal conductivity in graphene with different width. (**b**) Width-dependent thermal conductivity at $T$ = 300K. Solid circles are our experiment data in samples with 2.5μm in length, solid line is our fitting from Supplementary Equation (4); dash line and solid squares are from Bae *et al.*[34].

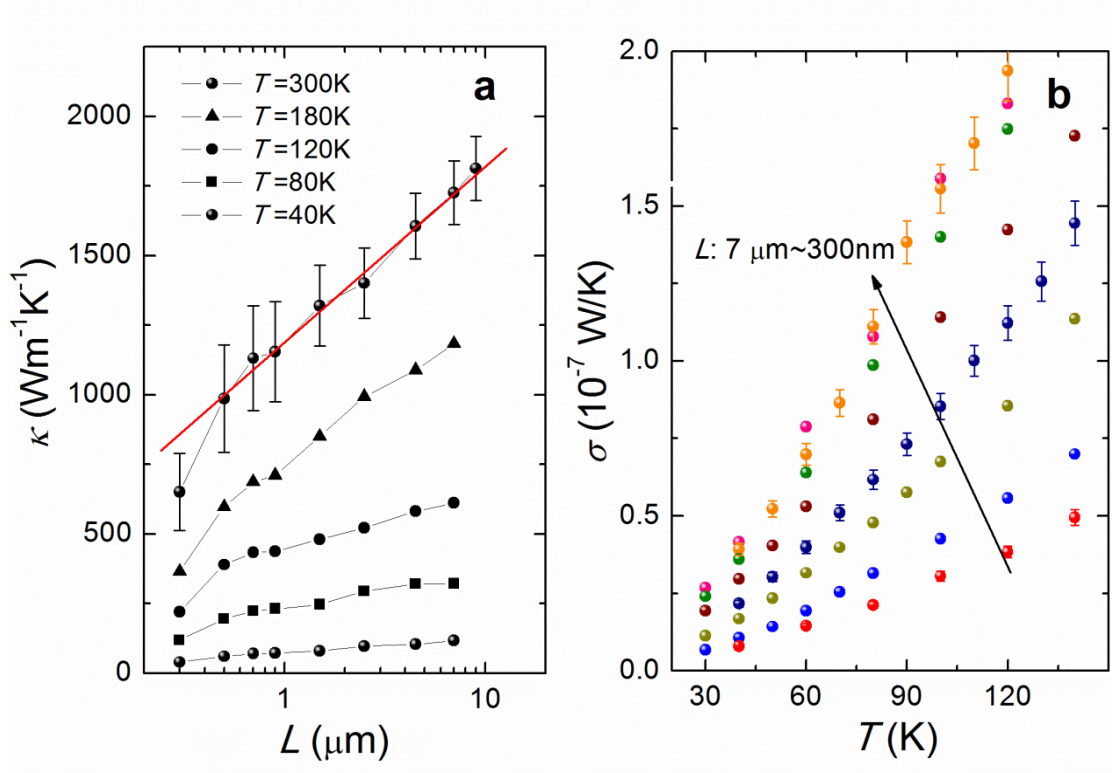

**Supplementary Figure 7 | Temperature-dependent thermal conduction.** (**a**) Thermal conductivity $\kappa$ vs. sample length at different temperatures with thermal contact resistance $R_c = 1.26 \times 10^6$ K/W. (**b**) Total measured thermal conductance $\sigma$ at low temperature.

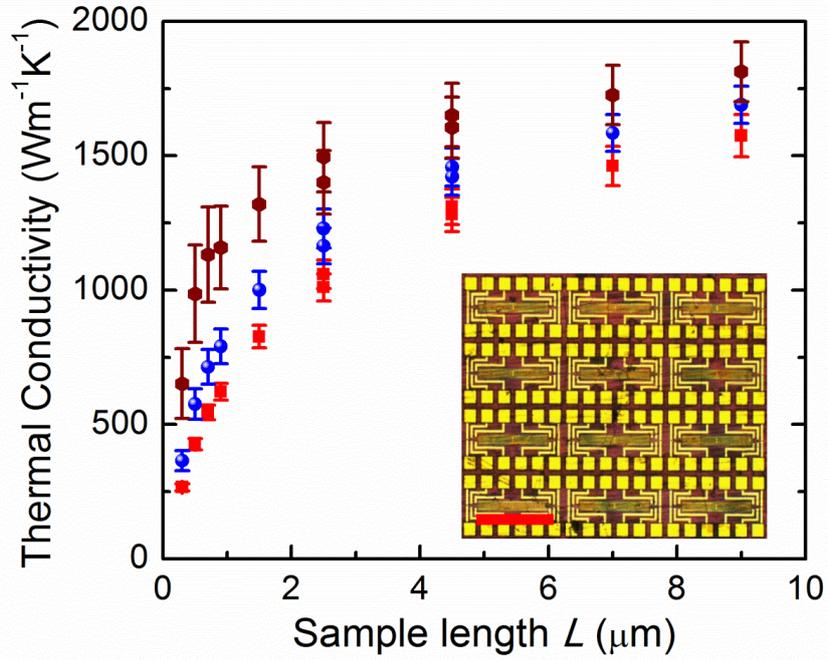

**Supplementary Figure 8 | Length dependence of $\kappa$.** The red squares, solid circles and solid diamonds are the extracted intrinsic thermal conductivity when assuming $R_c$ contributes negligible, 5% ($5.5 \times 10^5$ K/W) and 11.5% ($1.26 \times 10^6$ K/W) to the total measured thermal resistance in 9μm-long sample. Also see Figure 3a. Insert: 3 × 4 device array for length-dependent measurements, the scale bar represents 1 mm.

**Supplementary Table 1 | Measurement uncertainty.**

| Sample length | $L$ : 300nm | | $L$ : 500nm | | $L$: 7µm | | $L$: 9µm | |
|---|---|---|---|---|---|---|---|---|
| temperature | $T = 30$K | $T = 300$K | $T = 40$K | $T = 300$K | $T = 40$K | $T = 300$K | $T = 30$K | $T = 300$K |
| $\Delta T_s$ | 2.84 K | 4.02 K | 1.4K | 4.88K | 0.73 K | 4.95K | N.A. | 4.28 K |
| uncertainty | 1.3 % | 0.9% | 2.5% | 0.74% | 4.9% | 0.7% | N.A. | 0.8% |

**Supplementary Table 2 | Total Error bar when considering both 5% measurement uncertainty (Supplementary Table 1) and 10% variation of the thermal contact resistance from different samples.**

| When concerning $R_c$=5.5 $\times 10^5$ K/W | | | | | | |
|---|---|---|---|---|---|---|
| $L$ (µm) | 0.3 | 0.5 | 0.7 | 0.9 | 1.5 | 2.5 |
| Error bar (Wm$^{-1}$K$^{-1}$) | 44.1 | 67.2 | 775 | 78.9 | 87.8 | 94.4 |
| $L$ (µm) | 2.5 | 4.5 | 4.5 | 7 | 9 | |
| Error bar (Wm$^{-1}$K$^{-1}$) | 87.7 | 97.9 | 94.9 | 97.7 | 100.5 | |

| When concerning $R_c$=1.26 $\times 10^6$ K/W | | | | | | |
|---|---|---|---|---|---|---|
| $L$ (µm) | 0.3 | 0.5 | 0.7 | 0.9 | 1.5 | 2.5 |
| Error bar (Wm$^{-1}$K$^{-1}$) | 129.9 | 181.3 | 177.3 | 153.9 | 138.3 | 129 |
| $L$ (µm) | 2.5 | 4.5 | 4.5 | 7 | 9 | |
| Error bar (Wm$^{-1}$K$^{-1}$) | 117.5 | 118.2 | 113.8 | 110.6 | 111.1 | |

**Supplementary Table 3 | Comparison of the results with that from other works.**

|  | Our result (CVD) | CVD [21,26,31] | Exfoliated [20,32,33] | Exfoliated [22] |
|---|---|---|---|---|
| $\kappa$ at RT ($Wm^{-1}K^{-1}$) | 1689±100 to 1813±111 | 2500+1100/-1050 to 3100±1000 | ~1800 to ~5300 | ~650 |
| Layers | Monolayer | Monolayer | Monolayer | Bilayer |
| Method | Thermal bridge | Raman | Raman | Thermal bridge |
| Geometry | Rectangle | Corbino | Corbino & Rectangle | Rectangle |
| Temperature | 300K | ~350K | ~350K | 300K |

**Supplementary Table 4 | The measured thermal interfacial resistance from different groups and the calculated contribution from thermal contact resistance in our work.**

|  | APL 95, 161901 [49] | Nano Lett 10, 1645 [21] | Nano Lett 10, 4363 [50] | Nano Lett 11,1195 [22] | Our results |
|---|---|---|---|---|---|
| Interface materials | G/SiO$_2$ | G/Au | Au/Ti/G/SiO$_2$ | G/Pt/SiN$_x$ | G/Pt/SiN$_x$ |
| $R_{interface}$ (m$^2$KW$^{-1}$) | (4.2-12)×10$^{-9}$ | (2.3-5.3)×10$^{-8}$ | 4×10$^{-8}$ | Assume the same with Ref [21,49,50] | |
| Contribution from contact resistance | N.A. | N.A. | N.A. | 1.6%-5.9% | 0.9-4% |

## Supplementary Note 1 | Error Bars

The error bars are determined by:

1. Measurement uncertainty (~ 5%)

We used the same experiment setup with P. Kim and L. Shi, in which the first thermal conductivity of single carbon nanotube was measured[46,47]. This measurement setup has been tested as one of the most successful techniques in measuring thermal transport properties of single nanowire, nanotube and free standing 2D materials. In this technique, the temperature uncertainty in Sensor membrane $\delta(\Delta T_s)$ has been estimated to be around 36mK when assuming a 1 percent gain accuracy of the lock-in amplifier[46]. Therefore, the measurement uncertainty is between 0.7% to 4.9% in all the samples we measured (see Supplementary Table 1). Of course there should be some uncertainty coming from the temperature fluctuation of the base temperature in the low temperature measurement system (Cryogenic 9T system). For simplicity, we assume 5% measurement uncertainty in all the samples.

2. Thermal contact resistance variations between samples

The data are collected from samples with the same batch of CVD growth and fabricated at the same time to reduce the contact resistance variation between samples. However, there should be some variation in $R_c$ between samples. We assume this number is around ±10% and consider this as error bars. Combining the 5% measurement uncertainty mentioned above, we show the overall error bars in Supplementary Table 2.

**Supplementary Note 2 | Comparison with the results from other groups**

We compare the room temperature thermal conductivity in our longest sample ($\kappa = 1689 \pm 100$ Wm$^{-1}$K$^{-1}$ with $R_c = 5.5 \times 10^5$ K/W or $\kappa = 1813 \pm 111$ Wm$^{-1}$K$^{-1}$ with $R_c = 1.26 \times 10^6$ K/W, respectively) with previously reported experimental results (Supplementary Table 3). The value for $\kappa$ reaches the lower bound of that based on the Raman measurements, which show average values ranging from ~2500 +1100/-1050 Wm$^{-1}$K$^{-1}$ to ~ 3100 $\pm$ 1000 Wm$^{-1}$K$^{-1}$ in CVD SLG[21,26,31] and ~ 1800 Wm$^{-1}$K$^{-1}$ to ~ 5300 Wm$^{-1}$K$^{-1}$ in exfoliated graphene[20,32,33] (Since measured thermal conductivity increases with sample length, we believe thermal conductivity will exceed 2000 Wm$^{-1}$K$^{-1}$ if sample is large enough. This will be comparable or larger than that in graphite, which however is challenging in this experiment). This is in reasonably agreement considering the uncertainty of Raman measurements (~40%)[52] and differences in sample geometry and size[20,21,26,31,32]. On the other hand our results show a thermal conductivity which is more than two times larger than that obtained using the same measurement technique in suspended exfoliated bilayer graphene[22]. This is expected due to the lack of interlayer coupling in SLG[32].

## Supplementary Note 3 | Thermal contact resistance

To be consistent and eliminate sample-to-sample variations in contact resistance, the data were collected from the same batch of CVD-grown graphene and from two different $SiN_x$ wafers, each of them integrated with $6 \times 7$ devices array with different length and fabricated at the same time.

General speaking, total measured resistance $R_{total}$ for our devices is

$$R_{total} = R_c + R_{graphene} = R_c + \frac{L}{A\kappa} \quad\ldots\ldots\ldots(1)$$

a linear extrapolation of $R_{total} \sim L$ will gives an estimated value of $R_c$. However, it is important to note that this linear extrapolation is only valid in diffusive regime where $\kappa$ is length independent; since the length dependence of $\kappa$ is apriori unknown based on the measured thermal resistance data alone, we use other three approaches to calculate the thermal contact resistance:

**(1).** The thermal contact resistance $R_c$ can be calculated using thermal interfacial resistance ($R_{interface}$). The two have been shown to be related as[22]:

$$R_c/2 = [\sqrt{\frac{\kappa A w}{R_{interface}}} \tanh(\sqrt{\frac{w}{\kappa A R_{interface}}} l_c)]^{-1} \quad\ldots\ldots(2)$$

Supplementary Equation (2) takes into account the inherently larger contact area in 2D systems and the heat spreading under the contacts, where $R_c$ is thermal contact resistance, the number 2 in $R_c/2$ indicate two contacts on the two sides of graphene sample, $\kappa$ is thermal conductivity of supported graphene, $A$ is the contact area between sample and electrode, $w$ is the sample width, $l_c$ is the contact length, and $R_{interface}$ is the thermal interfacial resistance.

The electrical contact resistance of the samples is within $\sim 1$ k$\Omega$, indicating clean interfaces between graphene and Pt electrodes. The presence of Cr/Au top contacts (top red contacts in Supplementary Fig. 5d) further improves the thermal contact. Hence we assume that room temperature thermal interfacial resistance of Graphene/Pt/$SiN_x$ interface in our devices is comparable to $(4.2-12) \times 10^{-9}$ m$^2$KW$^{-1}$ in graphene/$SiO_2$ interface[49], $(2.3-5.3) \times 10^{-8}$ m$^2$KW$^{-1}$ in graphene/Au interface[21], $4 \times 10^{-8}$ m$^2$KW$^{-1}$ in Au/Ti/graphene/$SiO_2$ interface[50] (see Supplementary Table 4). From this, $R_c$ in our samples is calculated to be $\sim 9.9 \times 10^4$ K/W to $\sim 4.4 \times 10^5$ K/W, which contribute $\sim 0.9\%$ to $\sim 4\%$ of the total measured thermal resistance in sample with $L = 9$ μm at room temperature. Please note that this calculated thermal contact resistance ($\leq 4\%$) does not account for the interface between suspending graphene and the metal-clamped graphene.

**(2).** It is challenge to extract different thermal resistance component in the suspended graphene, especially in the submicron samples where phonons probably in the ballistic-diffusive regime. However, Prasher[51] has suggested that thermal contact resistance can be calculated from a simple model: $R_{total} = R_c + R_{ballistic} + R_{diff}$, where $R_{ballistic}$ is from ballistic phonons and $R_{diff}$ is from diffusive component. Following Bae *et al.*[34], the crossover of 2D thermal conductivity from ballistic to diffusive can be written through this model with:

$$R_{ballistic} = \frac{1}{\sigma_{ballistic}}, \quad R_{diff} = \frac{2L}{\pi\lambda}\frac{1}{\sigma_{ballistic}}, \quad \sigma_{ballistic}/A \approx \left[\frac{1}{4.4\times 10^5 T^{1.68}} + 1/(1.2\times 10^{10})\right]^{-1} \text{Wm}^{-2}\text{K}^{-1},$$

where *A* is cross section area, *L* is sample length and *λ* is phonon mean free path. We use this simple model to fit the data from samples with length ranging from 300nm to 1.5 μm. $R_c$ is fitted to be around ~$1.26 \times 10^6$ K/W and contribute around 11.5% of the total thermal resistance in 9μm - sample.

**(3).** We assume that thermal conductivity follow ~ log *L* behavior and directly fit $R_{total} = R_c + L/(\kappa A)$, where $\kappa$ ~ log *L* and extract $R_c$ to be ~$7.6 \times 10^5$ K/W, which contribute around 6.9% of the total thermal resistance in 9μm-sample.

It is important to note that whether phonons are in the ballistic-diffusive crossover or follows ~log *L* divergence is apriori unknown and we are not sure which approach provide the best thermal contact resistance $R_c$ in our measurement. However, we can safely assume $R_c$ is some value ranging from ~ $9.9 \times 10^4$ K/W to ~ $1.26 \times 10^6$ K/W (which contribute around 0.9% to 11.5% of $R_{total}$ in the sample with 9 μm length) and plot $\kappa$ ~ log *L* with different $R_c$ (negligible, $5.5 \times 10^5$ K/W and $1.26 \times 10^6$ K/W), as shown in Figure 3a.

## Supplementary Note 4 | Simulation

**Linear response regime:** In order to ensure that our simulations are within the linear response regime, we present in the top panel of Supplementary Fig. 1, the relationship between heat flux and temperature gradient obtained from RNEMD simulations. We observe a linear relationship between heat flux and temperature gradient up to gradients of approximately 2 K/nm. Furthermore, we observe that the thermal conductivity approaches a constant value as the gradient decreases, and that this value is in good agreement with the NEMD result.

**Non-equilibrium phonon populations:** In order to characterize the effects of the non-equilibrium dynamics in the behavior of phonons, we compare equilibrium and non-equilibrium phonon populations directly by means of the vibrational density of states (VDOS). The VDOS of each slab in the simulation cell is calculated from the Fourier transform of the velocity autocorrelation function of the atoms inside the respective slab. The VDOS can be resolved in to two in-plane directions (LA, LO, TA, TO) and one out-of-plane (ZA, ZO). An increase or decrease of the population at each frequency is given by a coefficient defined as:

$$\Delta n = \frac{1 + \int d\omega D_{NEMD}(\omega)}{1 + \int d\omega D_{EMD}(\omega)} - 1 \quad\quad\quad (3)$$

We take a simulation cell with length $L = 170.5$ nm (70,400 atoms) and divide it in 20 slabs (see Supplementary Fig. 2). The cold slab is labeled 1 and the hot slab is labeled 11. Close to the cold and hot layers, we observe an increase in population of out-of-plane modes and a reduction in the population of longitudinal modes, which compensate each other almost exactly. The population of transverse modes and the total population in each slab is essentially unchanged in NEMD and EMD.

## Supplementary Note 5 | Width-dependent thermal conductivity

In order to study the length-dependent thermal conductivity in graphene, it is important to fix the sample width as the two edges may also affect thermal conductivity by scattering the phonons, especially the phonons with long mean free path. Supplementary Fig. 6a shows thermal conductivity in 2.5μm-long samples with different width (negligible $R_c$). The measured thermal conductivity changes from 1054 Wm$^{-1}$K$^{-1}$ to 1186 Wm$^{-1}$K$^{-1}$ when width changing from 1.5 μm to 4 μm, demonstrating a weak width-dependent thermal conductivity in this range of width.

Bae *et al.* [34] used a simple empirical model:

$$\kappa(w,L) \approx \left[ \frac{1}{c}\left(\frac{\Delta}{w}\right)^n + \frac{1}{\kappa(L)} \right]^{-1} \quad\quad\quad\quad(4)$$

to understand the width-dependent data in supported graphene when considering that $\kappa$ is limited by phonon scattering with edge disorder. Here $\Delta$ is the root-mean-square edge roughness, $c$ is fitting parameter. The solid line in Supplementary Fig. 6b is our best fitting when using $\Delta$ = 0.6nm, and $c$ = 0.04 Wm$^{-1}$K$^{-1}$ (same parameters with Bae *et. al* [34]) at $T$ = 300K. The steeper width dependence in our fitting is probably due to the fact that our graphene is suspended and we have much larger length ($L$ = 2.5μm) than that in Bae *et. al* 's work ($L$ = 250nm)[34]. This is also consistent with previous theoretical work that graphene-substrate coupling affect size dependence of thermal conductivity in supported graphene[35].

## Supplementary References: